\documentclass[10pt,conference]{IEEEtran}
\IEEEoverridecommandlockouts
\usepackage{fancyhdr}
\usepackage{cite}
\usepackage{makecell}
\usepackage{amsmath,amssymb,amsfonts}
\usepackage{algorithmic}
\usepackage{graphicx}
\usepackage{textcomp}
\usepackage{xcolor}
\usepackage{subfigure}
\usepackage{graphicx}
\usepackage{float}
\usepackage{subfig}
\PassOptionsToPackage{subfigure}{tocloft}
\usepackage{CJK}
\usepackage[font={small}]{caption}
\usepackage{amsmath,amssymb,amsfonts}
\usepackage{graphicx}
\usepackage{textcomp}
\usepackage{xcolor}
\usepackage{amsmath}
\usepackage{amssymb}
\usepackage{graphicx}
\usepackage{epstopdf}
\usepackage{multirow}
\usepackage{stfloats}
\usepackage{cite}
\usepackage{color}
\usepackage[normalem]{ulem}
\usepackage{enumerate}
\usepackage{bbm}
\usepackage[top= 0.75in,bottom = 1.02in ,left= 0.625in,right= 0.625in]{geometry}
\usepackage{times,amsmath,amsthm,color,amssymb,graphicx,epsfig,algorithm,algorithmic,bm,multirow,cite}

\newtheorem{remark}{Remark}

\newtheorem{myDef}{Definition}

\newtheorem{myOpt}{Optimization}

\begin{document}

\title{Collaborative Deep Reinforcement Learning for Resource Optimization in Non-Terrestrial Networks}

\author{\IEEEauthorblockN{Yang Cao$^\text{1,2}$, Shao-Yu Lien$^\text{3}$, Ying-Chang Liang$^{\text{1},\text{2}}$, {\it{Fellow, IEEE}},\\ Dusit Niyato$^\text{4}$, {\it{Fellow, IEEE}}, and Xuemin (Sherman) Shen$^\text{5}$, {\it{Fellow, IEEE}}}
\IEEEauthorblockA{\small $^\text{1}$Yangtze Delta Region Institute (Huzhou), University of Electronic Science and Technology of China, Huzhou, P. R. China}
\IEEEauthorblockA{\small $^\text{2}$University of Electronic Science and Technology of China, Chengdu, P. R. China}
\IEEEauthorblockA{\small $^\text{3}$National Yang Ming Chiao Tung University, Tainan City, Taiwan}
\IEEEauthorblockA{\small $^\text{4}$Nanyang Technological University, Singapore}
\IEEEauthorblockA{\small $^\text{5}$University of Waterloo, Waterloo, Canada}}

\maketitle

\thispagestyle{fancy}
\protect\lhead{\small{© 2023 IEEE. Personal use of this material is permitted.  Permission from IEEE must be obtained for all other uses, in any current or future media, including reprinting/republishing this material for advertising or promotional purposes, creating new collective works, for resale or redistribution to servers or lists, or reuse of any copyrighted component of this work in other works.}
\small{DOI: 10.1109/PIMRC56721.2023.10294047}}
\renewcommand{\headrulewidth}{0pt}

\begin{abstract}
Non-terrestrial networks (NTNs) with low-earth orbit (LEO) satellites have been regarded as promising remedies to support global ubiquitous wireless services. Due to the rapid mobility of LEO satellite, inter-beam/satellite handovers happen frequently for a specific user equipment (UE). To tackle this issue, earth-fixed cell scenarios have been under studied, in which the LEO satellite adjusts its beam direction towards a fixed area within its dwell duration, to maintain stable transmission performance for the UE. Therefore, it is required that the LEO satellite performs real-time resource allocation, which however is unaffordable by the LEO satellite with limited computing capability. To address this issue, in this paper, we propose a two-time-scale collaborative deep reinforcement learning (DRL) scheme for beam management and resource allocation in NTNs, in which LEO satellite and UE with different control cycles update their decision-making policies through a sequential manner. Specifically, UE updates its policy subject to improving the value functions of both the agents. Furthermore, the LEO satellite only makes decisions through finite-step rollouts with a reference decision trajectory received from the UE. Simulation results show that the proposed scheme can effectively balance the throughput performance and computational complexity over traditional greedy-searching schemes.
\end{abstract}


\section{Introduction}
Targeting at constructing high-throughput global telecommunication systems, low-earth-orbit (LEO) satellites with low propagation attenuation levels have become key enabling paradigms in non-terrestrial networks (NTNs) \cite{Su2019}. However, the rapid moving speed around the earth renders the LEO satellite suffering from dramatic variations on channel condition and resource availability. In this case, how to maintain stable transmission performance is a critical concern in LEO-enabled NTNs. Although the transmission consistency for a specific ground user equipment (UE) can be handled through seamless constellation design, inter-beam/satellite handovers may happen frequently. This indicates that an extremely large amount of signaling overheads would be incurred, which are far beyond those of the terrestrial networks.

To address the above issue, the earth-fixed cell scenario has been widely investigated in the third generation partnership project (3GPP) \cite{3GPP2021}, in which the transmitting beam is kept to point toward the target UE within the dwell duration. In this case, the LEO satellite is required to act as base station to perform frequency bandwidth, transmit power, time slot and beam direction configurations timely. In \cite{Palacios2021}, the transmitting beam design was investigated based on the relative movement relationship between the LEO satellite and the ground UE. Additionally, a distributed precoding method for satellite swarm was proposed in \cite{Roper2022} to maximize the sum throughput at the ground gateway. Although these geometric model based schemes can bring competitive performances when line-of-sight (LOS) links exist, the LOS link may  in mountain or forest areas, and the scattering on landform may lead to a huge amount of fast-scale variations, i.e., the doppler shifts and multi-path effects. To fully capture these fast-scale variations, deep neural networks (DNNs) based data-driven schemes have been adopted to enhance the throughput performance of NTN. For instance, deep reinforcement learning (DRL) was adopted to construct intelligent beam-hopping technique for LEO satellites \cite{Liao2020}.

Nevertheless, in practical NTN deployment scenarios, the computing capability of the LEO satellite is normally constrained subject to the low payload sizes and low-efficiency energy supplies, and resource optimization tasks involving various factors cannot be completed by the LEO satellite within a short dwell duration. Consequently, it is in an urgent requirement to alleviate the computing burdens of the LEO satellite, and two candidate schemes are considered, i.e.,  utilizing cross-layer computing capability and decreasing the decision-making frequency. Specifically, in the former scheme, the powerful computing capabilities of ground UEs are utilized to address resource optimizations for the LEO satellite. For example, multiple ground UEs can be scheduled to maintain a global optimization model for the LEO satellite \cite{Razmi2022}. While in the latter scheme, the periodicity of the satellite ephemeris is utilized to decrease the decision-making frequency of the LEO satellite. In \cite{Zhao2021}, a periodic beam codebook design was proposed to balance the complexity and throughput performance. However, the throughput performance may degrade sharply if variations on channel gains and traffic loads within the large control cycle are not considered.

Toward further tackling the tradeoff between the computing complexity and the throughput performance, a combination of the above two candidates is considered in our work. Specifically, the high-dimensional optimization problem at the LEO satellite side is first decomposed into coupled sub-problems performed by the LEO satellite and UE in the NTN. Targeting at decreasing the computing burdens, the LEO satellite makes beam/resource management decisions valid in a large control cycles, i.e., at a level of hundreds of milliseconds. In contrast, UE with powerful computing capability adjusts its beam/resource strategy with a small control cycle, i.e., at a level of milliseconds, to further improve its throughput performance. Due to the differences of control cycles, the ground UE should be responsible for most computations in this scheme. To practically implement the above functions, in this paper, we propose a two-time-scale collaborative DRL scheme for beam management and resource allocation in NTNs, in which the LEO satellite and UE act as space-tier and ground-tier agent perform decision-making tasks with different control cycles to solve their individual \emph{Markov decision processes} (MDPs). Finally, comprehensive simulations are conducted to show the effectiveness of the proposed scheme in balancing the transmission performance and computational complexity.
\section{System Model and Problem Formulation}

\begin{figure}
\centering
\includegraphics[scale = 0.7]{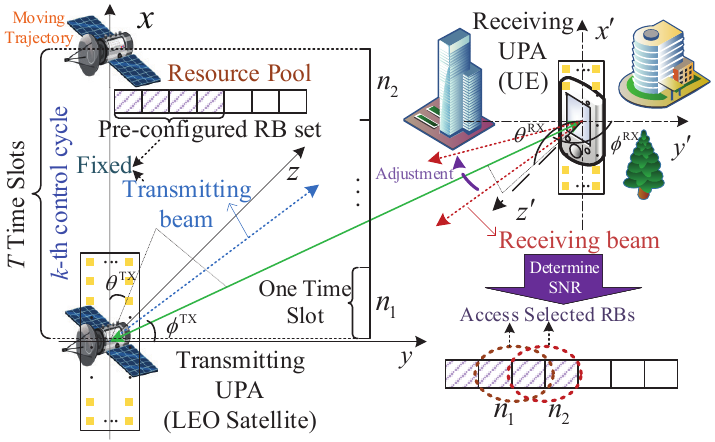}
\caption{The LEO downlink transmission model considered in this paper.}\label{fig:systemodel}
\end{figure}

\subsection{System Model}
As illustrated in Fig. \ref{fig:systemodel}, an LEO downlink transmission scenario is considered in this paper, in which an LEO satellite moving along the pre-designed orbit services a ground UE with the earth-fixed cell scenario. To this end, uniform planar arrays (UPAs) with $N_t = N_t^x \times N_t^y$ and $N_r = N_r^{x'} \times N_r^{y'}$ antennas are adopted at the LEO satellite side and UE side, respectively, where $N_t^x$ and $N_t^y$ ($N_r^{x'}$ and $N_r^{y'}$) denote the numbers of antennas at x- and y-axis ($x'$- and $y'$-axis) of the UPA at the LEO satellite (UE) side. Additionally, the \emph{orthogonal frequency division multiple access} (OFDMA) mechanism is adopted in this NTN model, and a resource pool ${\cal{M}}$ is composed of $M$ orthogonal resource blocks (RBs), i.e., ${\cal{M}} = \{0 , \ldots, M - 1\}$ . Particularly, multiple RBs are allowed to be allocated to the UE at a single time slot. To indicate whether RB $m$ is allocated to UE at time slot $n$, a binary indicator $b_{n,m}$ is adopted, i.e., $b_{n,m} = 1$ if RB $m$ is allocated to UE at time slot $n$, and $b_{n, m} = 0$ otherwise.

\subsubsection{Channel Gain Model}
According to \cite{Li2022}, the downlink (small-scale) channel gain between the LEO satellite and UE is mainly composed of four components, i.e., the doppler shift, propagation delay, transmitting and receiving array 3D-steering vectors,
\begin{align}\label{eq:channel}
{\bm{H}}_{n,m}  = \sum\limits_{l=0}^{L-1}\alpha_l e^{j2\pi[nT_sv_l-\frac{m}{T_s}\tau_l]}{\bm{a}}_r(\theta_l^{\text{Rx}}, \phi_l^{\text{Rx}}){\bm{a}}_t^{\text{H}}(\theta_l^{\text{Tx}}, \phi_l^{\text{Tx}}),
\end{align}
where $T_s$ is the time duration of one OFDM symbol, $L$ is the number of multi-paths, $\alpha_l$, $v_l$ and $\tau_l$ are the complex-valued gain, Doppler shift and propagation delay at path $l$, and $(\cdot)^{\text{H}}$ is the Hermitian transpose operation. Besides, ${\bm{a}}_t(\theta_l^{\text{Tx}}, \phi_l^{\text{Tx}})$ and ${\bm{a}}_r(\theta_l^{\text{Rx}}, \phi_l^{\text{Rx}})$ are array 3D-steering vectors at path $l$ with angles at $(\theta_l^{\text{Tx}}, \phi_l^{\text{Tx}})$ and $(\theta_l^{\text{Rx}}, \phi_l^{\text{Rx}})$, respectively, i.e.,
\begin{align}
{\bm{a}}_t(\theta_l^{\text{Tx}}, \phi_l^{\text{Tx}}) = {\bm{a}}_{t,x}(\theta_l^{\text{Tx}}, \phi_l^{\text{Tx}}) \otimes {\bm{a}}_{t,y}(\theta_l^{\text{Tx}}, \phi_l^{\text{Tx}}),\\
{\bm{a}}_r(\theta_l^{\text{Rx}}, \phi_l^{\text{Rx}}) = {\bm{a}}_{r,x'}(\theta_l^{\text{Rx}}, \phi_l^{\text{Rx}}) \otimes {\bm{a}}_{r,y'}(\theta_l^{\text{Rx}}, \phi_l^{\text{Rx}}),
\end{align}
where ${\bm{a}}_{t,x}(\theta_l^{\text{Tx}}, \phi_l^{\text{Tx}})$ and ${\bm{a}}_{t,y}(\theta_l^{\text{Tx}}, \phi_l^{\text{Tx}})$ are the steering vectors on the $x$- and $y$-directions of the transmitting UPA, which can be given as
\begin{align}\nonumber
{\bm{a}}_{t,x}(\theta_l^{\text{Tx}}, \phi_l^{\text{Tx}}) &= \frac{1}{\sqrt{N_{t}^{x}}}[1, e^{j\frac{2\pi}{\lambda}d_t\sin(\phi_l^{\text{Tx}})\cos(\theta_l^{\text{Tx}})}, \ldots,\\
& e^{j\frac{2\pi}{\lambda}(N_{t}^{x} - 1)d_t\sin(\phi_l^{\text{Tx}})\cos(\theta_l^{\text{Tx}})} ]^{\text{T}},\\\nonumber
{\bm{a}}_{t,y}(\theta_l^{\text{Tx}}, \phi_l^{\text{Tx}}) &= \frac{1}{\sqrt{N_{t}^{y}}}[1, e^{j\frac{2\pi}{\lambda}d_t\cos(\phi_l^{\text{Tx}})}, \ldots, \\
&e^{j\frac{2\pi}{\lambda}(N_{t}^{y} - 1)d_t\cos(\phi_l^{\text{Tx}})} ]^{\text{T}},
\end{align}
where $\lambda$ is the wave length, $(\cdot)^{\text{T}}$ is the transpose operation, and $d_t$ is the inter-antenna spacing of transmitting UPA. Similarly, with the inter-antenna spacing of $d_r$, the steering vectors  with angles of $(\theta_l^{\text{Rx}}, \phi_l^{\text{Rx}})$ on the $x'$- and $y'$- direction of the receiving UPA can be expressed as
\begin{align}\nonumber
{\bm{a}}_{r,x'}(\theta_l^{\text{Rx}}, \phi_l^{\text{Rx}}) &= \frac{1}{\sqrt{N_{r}^{x'}}}[1, e^{j\frac{2\pi}{\lambda}d_r\sin(\phi_l^{\text{Rx}})\cos(\theta_l^{\text{Rx}})}, \ldots,   \\
& e^{j\frac{2\pi}{\lambda}(N_{r}^{x'} - 1)d_r\sin(\phi_l^{\text{Rx}})\cos(\theta_l^{\text{Rx}})} ]^{\text{T}},\\
{\bm{a}}_{r,y'}(\theta_l^{\text{Rx}}, \phi_l^{\text{Rx}}) &= \frac{1}{\sqrt{N_{r}^{y'}}}[1, e^{j\frac{2\pi}{\lambda}d_r\cos(\phi_l^{\text{Rx}})}, \ldots,  \\
& e^{j\frac{2\pi}{\lambda}(N_{r}^{y'} - 1)d_r\cos(\phi_l^{\text{Rx}})} ]^{\text{T}}.
\end{align}

\subsubsection{Downlink Transmission Model}
Denote transmitting beam of angles $(\theta_t^n, \phi_t^n)$ at the LEO satellite and receiving beam of angles $(\theta_r^n, \phi_r^n)$ at the UE as ${\bm{w}}_t(\theta_t^n, \phi_t^n) \in \mathbb{C}^{N_t \times 1}$ and ${\bm{w}}_r(\theta_r^n, \phi_r^n) \in \mathbb{C}^{N_r \times 1}$, respectively, which have the same definition with ${\bm{a}}_t(\cdot)$ and ${\bm{a}}_r(\cdot)$. The signal-to-noise-ratio (SNR) of UE in RB $m$ at time slot $n$ can be given as
\begin{align}\label{eq:snr}\nonumber
&\varpi_{n, m}(\theta_t^n, \phi_t^n, \theta_r^n, \phi_r^n)\\
 &= \frac{P_tL_n|{{\bm{w}}_r(\theta_r^n, \phi_r^n)}^{\text{H}}{\bm{H}}_{n, m}{\bm{w}}_t(\theta_t^n, \phi_t^n)|^2}{N_r\delta_z^2},
\end{align}
where $P_t$ is the transmit power, $L_n$ is the pathloss between the LEO satellite and UE at time slot $n$, ${{x}}_{n, m}$ is the transmit signal with unit power, and $\bm{z} \sim {\cal{CN}}({\bm{0}}, \delta_z^2{\bm{I}^{N_r \times N_r}})$ is the additive white Gaussian noise (AWGN) with the mean value of zero and variance of $\delta_z^2 = k_BT_NB$, where $k_B$, $T_N$, and $B$ are Boltzmann constant, noise temperature, and bandwidth of each RB, respectively. From \eqref{eq:snr}, it indicates that the SNR at the UE is mainly determined by the beam gain $|{{\bm{w}}_r}^{\text{H}}{\bm{H}}_{n,m}{\bm{w}}_t|^2$. Therefore, based on the Shannon capacity formula, the receiving rate of the UE in RB $m$ at time slot $n$ can be given by
\begin{align}
c_{n, m}(\theta_t^n, \phi_t^n, \theta_r^n, \phi_r^n) = B\log_2(1 + \varpi_{n, m}(\theta_t^n, \phi_t^n, \theta_r^n, \phi_r^n)).
\end{align}

\subsection{Problem Formulation}
In this paper, as shown in Fig. \ref{fig:systemodel}, a two-time-scale resource management scheme is adopted. Specifically, the LEO satellite is supposed to perform beam adjustment every $T$ time slots. In contrast, to further improve the SNR performance in each RB, the ground UE needs to adjust its receive beam at each time slot. In this case, the control cycle of the LEO satellite is $T$ time slots, and its control cycle index is $k = \lfloor{\frac{n}{T}}\rfloor$. The control cycle of the UE is one time slot, and its control cycle index is $i = n$. Particularly, the time slot lengths of the LEO satellite and UE are the same, and their time slot boundaries are aligned. On the other hand, since the limited size of resource pool in NTN, only a subset of RBs (i.e., ${\cal{{M}}}_{\text{LEO}}^n$) can be reserved for the UE. Furthermore, the UE should access RBs with respect to the highest SNR to satisfy its rate demand $D_{\text{UE}}^n$ at each time slot. Therefore, a long-term RB number minimization problem with respect to transmitting-receiving beam management and resource allocation can be mathematically formulated.

\begin{myOpt}In this paper, we aim at minimizing the average number of utilized RBs from time slot $0$ to $N-1$ to satisfy the dynamic UE receiving rate demand, i.e.,
\begin{align}
& \mathop{\min}\limits_{\{\theta_t^n, \phi_t^n, \theta_r^n, \phi_r^n\}, \{b_{n, m}\}, \{{\cal{{M}}}_{\text{LEO}}^n\} } \frac{1}{N}\sum\limits_{n = 0}^{N-1}\sum\limits_{m=0}^{M - 1}b_{n,m}\\\label{eq:cons_rate}
\mathrm {s.t.} & \sum\limits_{m \in {\cal{{M}}}_{\text{LEO}}^n}b_{n, m}c_{n, m}(\theta_t^n, \phi_t^n, \theta_r^n, \phi_r^n) \ge D_{\text{UE}}^n, \forall n,\\
& b_{n,m} \in \{0, 1\}, \forall (n,m),\\
& \theta_{\zeta}^n \in (0, \pi), \zeta \in \{t, r\}, \forall n,\\
& \phi_{\zeta}^n \in (0, \pi), \zeta \in \{t, r\}, \forall n,\\
& {\cal{{M}}}_{\text{LEO}}^n \subseteq   {\cal{{M}}}, \forall n.
\end{align}
\end{myOpt}

Due to the different control cycles of the LEO satellite and UE, an enormous number of combinations of involved variables should be considered to solve \textbf{Optimization 1}, which thus expands the decision space over that of the scenario with the same control cycle. However, even for a centralized controller, such an extremely high-dimensional optimizations cannot be solved in a real-time manner. On the other hand, although the whole optimization problems can be decomposed into sub-optimizations for different agents according to the time-scales of involved variables, it is still difficult to solve these decomposed optimizations with the lack of accurate variation patterns of the environment. To address the above issues, DRL is adopted to assist the LEO satellite and UE to learn variation patterns from continuous interactions from the environment. Additionally, since the limited computing capability of the LEO satellite cannot support capturing variations within a large-time-scale control cycle, UE with powerful computing capability should provide beneficial information to aid the LEO satellite to make effective decisions. To this end, we propose a two-time-scale collaborative DRL for beam management and resource allocation, in which the LEO satellite and UE perform individual decision-making tasks through a sequential manner. To alleviate the computing burdens of the LEO satellite, the UE updates its policies considering the mutual effects between policies of the UE and LEO satellite, and the LEO satellite just performs decision-making tasks based on the reference information provided by the UE.

\section{Proposed Two-time-scale Collaborative DRL Scheme}

\subsection{Two-time-scale MDP Formulation}

With the spirit of constructing multi-agent DRL scheme, \textbf{Optimization 1} is decomposed and formulated as individual MDPs for agents with different control cycles, i.e., ${\cal{M}}_H = <{\cal{S}}_H, {\cal{A}}_H, {\cal{P}}_H(\pi_H, \pi_L), {R}_H(\pi_H, \pi_L)>$ for LEO satellite and $ {\cal{M}}_L = <{\cal{S}}_L, {\cal{A}}_L, {\cal{P}}_L(\pi_H, \pi_L), {R}_L(\pi_H, \pi_L)>$ for UE, where ${\cal{S}}_H$, ${\cal{A}}_H$, $\pi_H$, ${\cal{P}}_H(\pi_H, \pi_L)$ and ${R}_H(\pi_H, \pi_L)$ (${\cal{S}}_L$, ${\cal{A}}_L$, $\pi_L$, ${\cal{P}}_L(\pi_H, \pi_L)$ and ${R}_L(\pi_H, \pi_L)$) are state space, action space, policy function, i.e., ${\bm{a}}_H = \pi_H({\bm{s}}_H)$ (${\bm{a}}_L = \pi_L({\bm{s}}_L)$), transition probability and reward function of higher-tier (lower-tier) agent, respectively. Particularly, the reward function and transition probability of each agent are influenced by the other agent's policies. For each agent, the objective is to learn the value function through optimizing the accumulated rewards from the environment, i.e., $V^{\pi_H}({\bm{s}}_H;\pi_L) = \mathbb{E}_{\pi_H}[\sum\limits_{n=0}^{\infty}{\gamma_H}^{n}R_H({\bm{s}}_H^n, \pi_H({\bm{s}}_H^n), \tau_L(\pi_L))|{\bm{s}}_H^0 = {\bm{s}}_H]$ or $V^{\pi_L}({\bm{s}}_L;\pi_H) = \mathbb{E}_{\pi_L}[\sum\limits_{n=0}^{\infty}{\gamma_L}^{n}R_L({\bm{s}}_L^n, \pi_L({\bm{s}}_L^n), \tau_H(\pi_H))|{\bm{s}}_L^0 = {\bm{s}}_L]$, where $\mathbb{E}[\cdot]$ is the expectation operator, and $\gamma_H$ and $\gamma_L$ are discount factors of higher-tier and lower-tier agents, respectively. With formulated MDPs, addressing \textbf{Optimization 1} is transformed as searching for the best policies $\pi_H^*$ and $\pi_L^*$ to maximize the sum of accumulated rewards of two agents, i.e.,
\begin{align}\nonumber
&\mathop{\max}\limits_{\pi_H, \pi_L} \mathbb{E}[\sum\limits_{k=0}^{N_H-1}[R_H({\bm{s}}_H^k, \pi_H({\bm{s}}_H^k), \tau_L(\pi_L)) +\\
&\sum\limits_{p=0}^{T-1}R_L({\bm{s}}_L^{Tk+p}, \pi_L({\bm{s}}_L^{Tk+p}), \tau_H(\pi_H))]].
\end{align}
where $N_H = \lfloor{\frac{N-1}{T}}\rfloor$ is the number of control cycles of the higher-tier agent at time slot $N-1$.

\subsection{Two-time-scale MDP Design}
Before introducing the details of the proposed collaborative DRL scheme, we first give the basic components of formulated MDPs of different agents in this section. The MDP of the LEO satellite with large-time-scale control cycle is presented below.
\begin{myDef} MDP of the LEO satellite
\begin{itemize}
\item{State Space:} Since the LEO satellite needs to capture the variation patterns on channel conditions, the state of the LEO satellite is composed of two groups: 1) position of the LEO satellite, i.e., $p_{\text{LEO}}^k = (x_{\text{LEO}}^k, y_{\text{LEO}}^k, z_{\text{LEO}}^k)$; 2) average SNR over selected RBs at each time slot in last control cycle, i.e.,  $\bar{\varpi}_{(k-1)T+p} = \frac{1}{M}\sum\limits_{m \in {\cal{M}}}\varpi_{(k-1)T+p, m}, p = 0, \ldots, T-1$,
\begin{equation}
{\bm{s}}_H^k = \{p_{\text{LEO}}^k; \bar{\varpi}_{(k-1)T},\\
 \ldots, \bar{\varpi}_{(k-1)T + T -1}\}.
\end{equation}
\item{Action Space:} Within each control duration, the LEO satellite needs to determine its transmitting beam direction and select a set of RB candidates, i.e.,
\begin{equation}
{\bm{a}}_H^k =\{\theta_t^k, \phi_t^k;  {\cal{{M}}}_{\text{LEO}}^k\}.
\end{equation}
\item{Reward Function:}According to constraint \eqref{eq:cons_rate} in \textbf{Optimization 1}, the LEO satellite should enhance the SNR in \eqref{eq:snr}, and then select RB sets with respect to the highest average receiving rate so that the number of RBs utilized by the UE can be reduced. Therefore, the reward function is defined as the average receiving rate within time slots satisfying constraint \eqref{eq:cons_rate} within each control cycle, i.e.,
\begin{align}\label{eq:h_reward}\nonumber
R_H^k= &\frac{1}{T}\sum\limits_{p = 0}^{T - 1}\mathbb{I}_{(\sum\limits_{m \in {\cal{{M}}}_{\text{LEO}}^k}b_{kT + p, m}c_{kT + p, m}  \ge D_{\text{UE}}^{kT + p})}(\\
&\frac{\sum\limits_{m \in {\cal{M}}_{\text{LEO}}^k}b_{kT + p, m}c_{kT + p, m}(\theta_t^k, \phi_t^k, \theta_r^{kT + p}, \phi_r^{kT + p})}{\sum\limits_{m \in {\cal{{M}}}_{\text{LEO}}^k}b_{kT + p, m}}),
\end{align}
where $\mathbb{I}_{(\cdot)}$ is the indicator function.
\end{itemize}

\rm{Subsequently, the MDP of the UE with small-time-scale control cycle is introduced in the following. }

\begin{myDef} MDP of the UE
\begin{itemize}
\item{State Space:} Toward utilizing variations on channel, there are two groups in the state of UE: 1) SNR in each RB at last time slot, i.e., \{$\varpi_{i, m}\}, \forall m \in {\cal{M}}$; 2) average signal strength at receive antennas over RBs at last time slot, i.e.,  ${\bm{y}}_r^{i-1} = \frac{1}{M}\sum\limits_{m \in \cal{M}}|{\bm{H}}_{i-1, m}{\bm{w}}_t(\theta_t^k, \phi_t^k)|$,
\begin{align}\nonumber
{\bm{s}}_L^i &= \{\varpi_{i-1, 0}, \ldots, \varpi_{i-1, M-1}; {\bm{y}}_r^{i-1}\}.
\end{align}
\item{Action Space:} The UE at each time slot needs to adjust its receiving beam direction and access proper RBs from pre-planned RB candidates, i.e.,
\begin{equation}
{\bm{a}}_L^i = \{\theta_r^i, \phi_r^i; b_{i,0}, \ldots, b_{i, M-1}\}.
\end{equation}
\item{Reward Function:}  To minimize the number of utilized RBs, the RB with respect to the highest receiving rate should be selected first to satisfy the constraint \eqref{eq:cons_rate}. To this end, the instantaneous reward function is composed of the average receiving rate of selected RBs and a satisfactory punishment to avoid wasting RBs, i.e.,
     \begin{equation}\label{eq:l_reward}
{\hat{R}}_L^i = \frac{\sum\limits_{m \in {\cal{M}}}b_{i,m}c_{i,m}(\theta_t^k, \phi_t^k, \theta_r^i, \phi_r^i)}{\sum\limits_{m \in {\cal{M}}}b_{i,m}} + \eta \Omega_{i},
\end{equation}
where $\Omega_{i} = \min[{\sum\limits_{m \in {\cal{M}}}b_{i,m}c_{i,m} - D_{\text{UE}}^{i}}, 0]$ is the satisfactory punishment, and $\eta$ is a punishment coefficient. Moreover, to alleviate fast variations in each time slot, a first-in-first-out (FIFO) buffer ${\cal{B}}_r$ is adopted to perform moving average over instantaneous rewards, and then the average value of ${\cal{B}}_r$ is adopted as the final reward function of the UE, i.e., $R_L^i = \mathbb{E}_{{\cal{B}}_r}[{\hat{R}}_L^i]$, where $\mathbb{E}_{{\cal{B}}_r}[\cdot]$ is the expectation over instantaneous rewards in buffer ${\cal{B}}_r$.
\end{itemize}
\end{myDef}

\end{myDef}

\subsection{Proposed Collaborative Scheme}
After presenting the main components of formulated MDPs, we provide the details of the proposed scheme in this section. Notice that the agents at different-tiers update their policy in a sequential manner \cite{Bertsekas2021}. Therefore, each policy updating stage is composed of two phases, i.e., the first phase for the lower-tier agent (i.e., the UE) and the second phase for the higher-tier agent (i.e., the LEO satellite). In the first phase, to improve both the value functions of two agents, the lower-tier agent should focus on not only the improvement of its policy on its value function, but also the effect of its policy to the higher-tier agent's value function. Therefore, the lower-tier agent should update its policy along the average direction toward maximizing the sum of value improvements of two agents with respect to the its policy, which therefore provide a ``correct'' improvement direction to the higher-tier agent in the second phase. To this end, toward obtaining stable improvements in value functions, the trust region policy optimization (TRPO) algorithm \cite{Schulman2015} in DRL is adopted by the lower-tier agent. Specifically, at the lower-tier agent, two DNNs with policy parameters $\bm{\varphi}_L$ and value function parameters ${\bm{\theta}}_L$ are deployed to approximate the policy and value function, respectively. Therefore, the loss function is calculated based on the product of the policy ratio $\frac{\pi_L({\bm{a}}_L^{n}|{\bm{s}}_L^{n}; \bm{\varphi}_L)}{\pi_L({\bm{a}}_L^{n}|{\bm{s}}_L^{n}; \bm{\varphi}^{'}_L)}$ and the sum of all the agents' estimated policy advantages, i.e.,
\begin{equation}\label{eq:actor_loss}
{\cal{L}}_{p} = \mathbb{E}[\frac{\pi_L({\bm{a}}_L^{n}|{\bm{s}}_L^{n}; \bm{\varphi}_L)}{\pi_L({\bm{a}}_L^{n}|{\bm{s}}_L^{n}; \bm{\varphi}^{'}_L)}(\tilde{A}_L({\bm{s}}_L^{n}, {\bm{a}}_L^{n})+\tilde{A}_H({\bm{s}}_H^{n}, {\bm{a}}_H^{n}))],
\end{equation}
where $\tilde{A}_L({\bm{s}}_L^{n}, {\bm{a}}_L^{n}) =  R_L({\bm{s}}_L^{n}, \pi_L({\bm{s}}_L^{n})) + \gamma_L V^{{\bm{\theta}}_L}({\bm{s}}_L^{n + 1})  - V^{{\bm{\theta}}_L}({\bm{s}}_L^{n})$  and $\tilde{A}_H({\bm{s}}_H^{n}, {\bm{a}}_H^{n})$ are the estimated policy advantages for higher-tier and lower-tier agents, respectively, and $\bm{\varphi}^{'}_L$ is the old policy parameters of the lower-tier agent. Moreover, to obtain accurate estimated policy advantages, the value function parameters ${\bm{\theta}}_L$ of the lower-tier agent are updated based on mean squared error (MSE) function and the accumulated rewards in its replay memory ${\cal{D}}_L$, i.e.,
\begin{equation}\label{eq:critic_loss}
{\cal{L}}_{v} = \mathbb{E}[(V^{{\bm{\theta}}_L}({\bm{s}}_L^{n}) - \sum\limits_{l = 0}^{T-n}{\gamma_L}^{l}R_L^{(n+l)})^2].
\end{equation}
Based on the updated policy $\bar{\pi}_L$ with parameters $\bar{\bm{\varphi}}_L$, the lower-tier agent generates a set of actions for future finite steps as a decision trajectory $\tau_L(\bar{\pi}_L)$, which is sent to the higher-tier agent as the reference information to further update its policy.

Next, in the second phase, with the predicted decision trajectory  $\tau_L(\bar{\pi}_L)$ from UE, the higher-tier agent improves its policy through rollout policy, i.e,
\begin{align}\label{eq:rollout}
\bar{\pi}_H &= \mathop{\arg\max}\limits_{\pi_H} \mathbb{E}[R_H({\bm{s}}_H^k, \pi_H({\bm{s}}_H^k); \tau_L(\bar{\pi}_L)) +\\\nonumber
        & \gamma_H V^{\pi_H}(h({\bm{s}}_H^k, \pi_H, \bar{\pi}_L); \bar{\pi}_L)],
\end{align}
where $h(\cdot)$ is the state transition function at space-tier, i.e., ${\bm{s}}_H^{k+1} = h({\bm{s}}_H^k, \pi_H, \pi_L)$.  Since there is an infinite horizon MDP at space-tier, the $\bar{n}$-rollout policy \cite{Bertsekas2021} according to $\pi_H$ can be adopted with respect to the assumption of ${\gamma_H}^{\bar{n}}V^{\pi_H}({\bm{s}}_H^{k + \bar{n}}; \bar{\pi}_L) \rightarrow 0$, i.e, $V^{\pi_H}({\bm{s}}_H^k; \bar{\pi}_L) \approx \sum\limits_{p= 0}^{\bar{n} - 1}{\gamma_H}^{p}R_H({\bm{s}}_H^{k+p}, \pi_H({\bm{s}}_H^{k+p}); \tau_L(\bar{\pi}_L))$,  to obtain an approximated value function at the higher-tier agent. Nevertheless, a large estimation step $\bar{n}$ is needed to obtain an accurate approximation value, but it may not be unaffordable for the LEO satellite with the limited computing capability. To this end, in this paper, we adopt one DNN with parameters ${\bm{\theta}}_H$ to estimate the tail value $V^{{\bm{\theta}}_H}({\bm{s}}_H^{n+\bar{n}})$ for the higher-tier agent. In this case, the approximated value at state ${\bm{s}}_H^k$ under updated ground-tier policy $\bar{\pi}_L$ can be rewritten as $V^{\pi_H}({\bm{s}}_H^n, \bar{\pi}_L) = \sum\limits_{p = 0}^{\bar{n} - 1}{\gamma_H}^{p}R_H({\bm{s}}_H^{k+p}, \pi_H({\bm{s}}_H^{k+p}); \tau_L(\bar{\pi}_L)) + {\gamma_H}^{\bar{n}}V^{{\bm{\theta}}_H}({\bm{s}}_H^{k+\bar{n}})$.
Similarly, the value function parameters ${\bm{\theta}}_H$ of the higher-tier agent are also updated through minimizing the MSE loss function defined in \eqref{eq:critic_loss}. Since the given decision trajectory is created according to the sum estimated advantages, the value function of the space-tier agent can also be improved monotonically over the original one.

\begin{remark}
The monotonic policy improvement can be achieved for either space-tier agent or ground-tier agent after each policy updating stage in the proposed collaborative DRL scheme.
\end{remark}
\begin{proof}
Please refer to Appendix A.
\end{proof}

Through iteratively performing the above policy updating procedures, both of the agents can gradually improve the value functions. In this case, the optimal policies maximizing the sum value functions can be obtained through a period of training. The complete procedure of the proposed scheme is illustrated in Algorithm 1.

\begin{algorithm}[thp]
\caption{Proposed Two-time-scale Collaborative DRL Scheme}
{\small
\begin{algorithmic}[1]
\STATE \textbf{Initialize Stage}:
\STATE Lower-tier agent constructs two DNNs with randomly initialized parameters, and a replay memory ${\cal{D}}_L$, and higher-tier agent constructs one DNN with randomly initialized parameters, and a replay memory ${\cal{D}}_H$.
\STATE \textbf{Training Stage}:
\REPEAT
\STATE Higher-tier agent sends its current state-action pair $({\bm{s}}_H^{k}, {\bm{a}}_H^{k})$ to the ground-tier agent.
\FOR {$i = 0$ to $T-1$}
\STATE Lower-tier agent obtains action ${\bm{a}}_L^i$ based on the current state and constructed DNN-based policy function.
\STATE Lower-tier agent executes the obtained action and receives an instantaneous reward from the environment.
\STATE Lower-tier agent stores obtained experience $\{ {\bm{s}}_L^i,{\bm{a}}_L^i, R_L^i\}$ into its replay memory ${\cal{D}}_L$.
\STATE Lower-tier agent samples experiences from ${\cal{D}}_L$ to calculate loss function based on \eqref{eq:actor_loss} and \eqref{eq:critic_loss}.
\STATE Lower-tier agent updates DNN parameters ${\bm{\varphi}}_L$ and ${\bm{\theta}}_L$ through minimizing the calculated losses.
\ENDFOR
\STATE Lower-tier agent generates future decision trajectory $\tau_L(\bar{\pi}_L)$ according to the updated policy.
\STATE Higher-tier agent receives reward and reference decision trajectory from the ground-tier agent and visits a new state ${\bm{s}}_H^{k+1}$.
\STATE Higher-tier agent obtains decision ${\bm{a}}_H^{k+1}$ through performing rollout policy in \eqref{eq:rollout} based on $\tau_L(\bar{\pi}_L)$.
\STATE Higher-tier agent stores the experience $\{ {\bm{s}}_H^{k},{\bm{a}}_H^{k}, R_H^{k}\}$ into its replay memory ${\cal{D}}_H$.
\STATE Higher-tier agent updates parameters ${\bm{\theta}}_H$ through minimizing the MSE of sampled experiences from ${\cal{D}}_H$.
\UNTIL{The parameter change at different tiers is less than a given error threshold $\epsilon$.}
\end{algorithmic}}
\end{algorithm}

\section{Performance and Evaluation}
\subsection{Simulation Settings}
To evaluate the performance of the proposed scheme, an LEO satellite with $K = 60$ RBs is deployed to service a ground UE at $(5045.27, 3881.81, -393.28)$ km in earth-centered-earth-fixed coordinate system. All the RBs are divided into $3$ distinct RB grpups, and the RB allocation procedure is implemented in the unit of RB group. The parameters of channel gain model are selected based on the 3GPP document \cite{3GPP2019}, and the LEO orbit is created through the Satellite Communication Box of MATLAB. Additionally, the minimum elevation angle for transmission is $\frac{\pi}{6}$, and the bandwidth of each RB is 180 kHz. The receiving rate demand is modeled with a Poisson distribution with $\lambda = 2$, and the unit value is $10$ megabits per slot. A non-codebook design is adopted for determining the transmitting/receiving beam directions, in which a discrete angle is added to the initial beam direction as the beam direction at the current time slot. The other main simulation parameters are summarized in Table I.

\addtolength{\topmargin}{0.04in}
\begin{table}\label{tab:2}
\caption{Parameters for simulation.}
\centering
\footnotesize
\begin{tabular}{llll}
\hline
  Parameters & Value \\
  \hline
  LEO transmission power & 30 dBW \\
  LEO/UE antenna gain  &  30 dBi \\
  Carrier frequency & 4 GHz (S band)\\
  Number of arriving demands ($N_{UE}$) & Poisson distribution with $\lambda = 2$\\
  Replay memory size of LEO/UE & 1200/9600\\
  Discounting factor of LEO/UE & 0.99\\
  Value function at LEO/UE & $400\times300\times200$\\
  Policy function at UE & $300\times200\times200$\\
  \hline
 \end{tabular}
 \vspace{-0.3cm}
\end{table}

For performance comparison, we adopt independent DRL scheme and proposed scheme without sum advantage estimation (i.e., single-estimation scheme). Additionally, traditional separated optimization schemes are also adopted as benchmarks. Specifically, brute-force searching (BFS) scheme and periodic beam update (PBU) scheme in \cite{Zhao2021} are adopted for beam management. For RB allocation, greedy and multi-armed bandit (MAB) schemes are utilized. Particularly, the BFS with greedy allocation scheme can achieve the upper bound on the receiving rate performance.

\subsection{Results and Analysis}
\begin{figure}[htb]
	\begin{minipage}[b]{0.49\linewidth}
        \centering
        \includegraphics[scale=0.16]{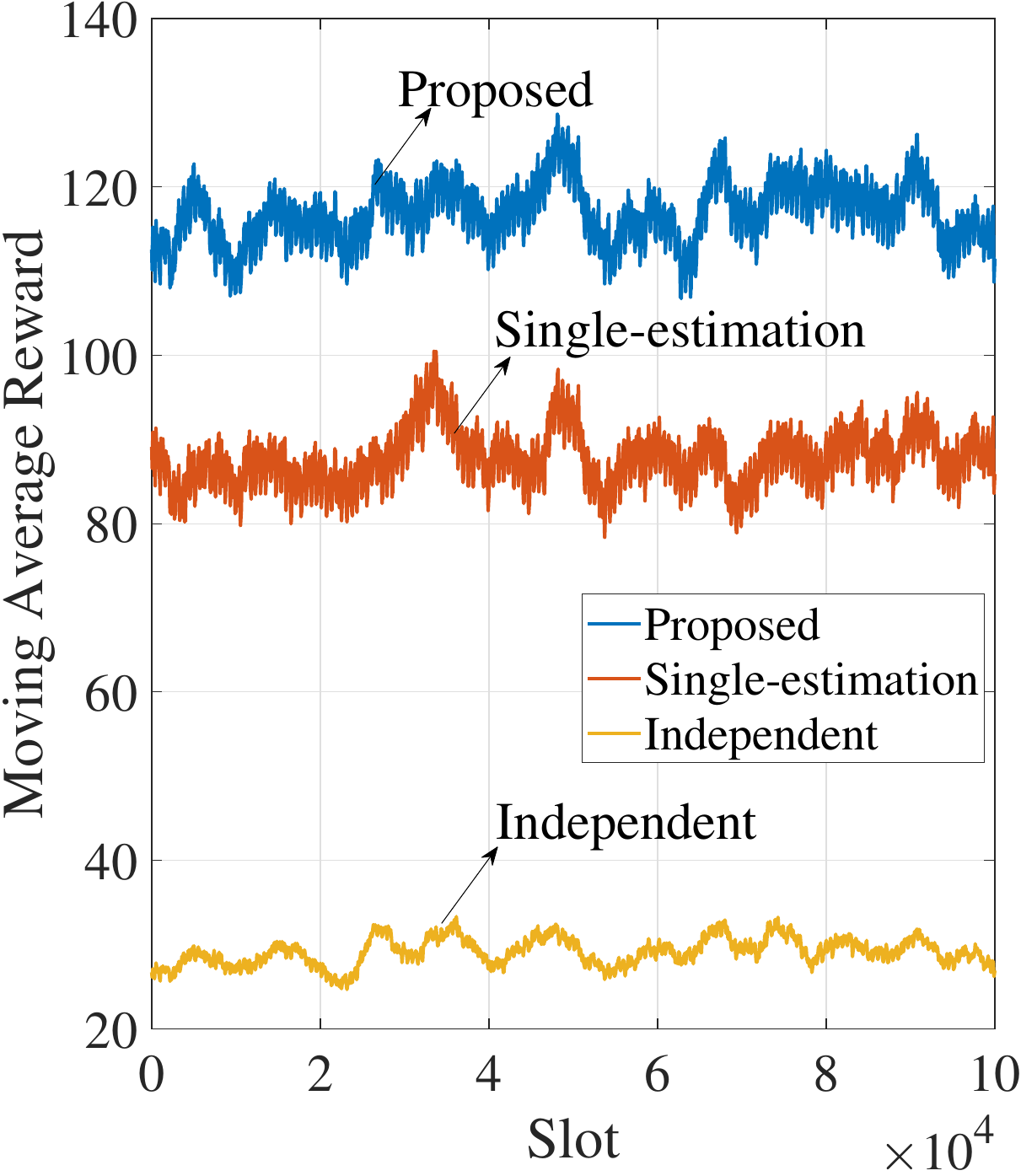}
    \end{minipage}%
    \begin{minipage}[b]{0.49\linewidth}
        \centering
        \includegraphics[scale=0.16]{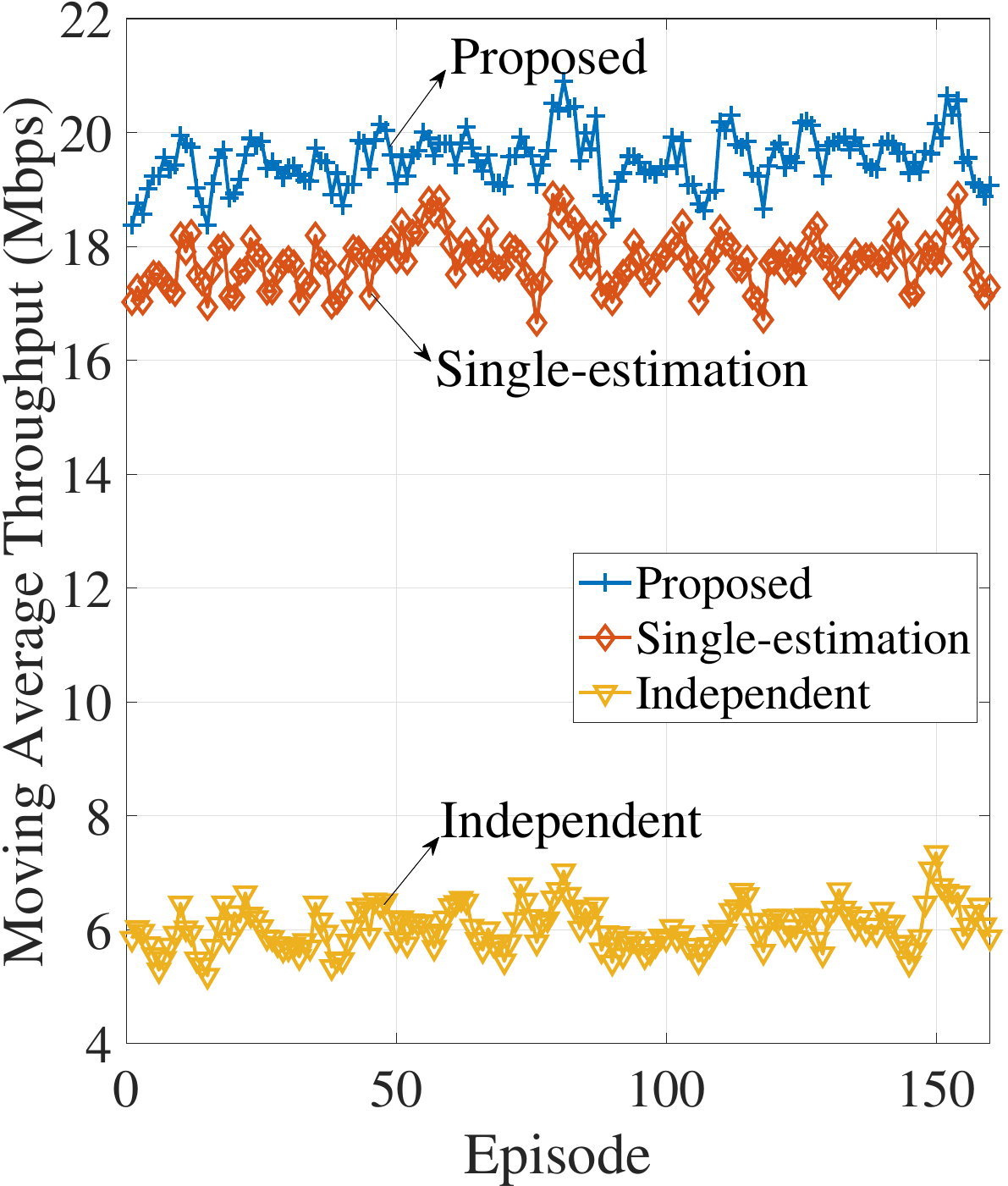}
    \end{minipage} \\[0.90mm]
\caption{Moving average reward and throughput with different DRL-based algorithms.}\label{fig:drl}
  \vspace{-0.5em}
\end{figure}
Fig. \ref{fig:drl} illustrates the moving average reward and throughput performance of the proposed scheme and other DRL-based schemes. One episode is defined as the service duration when the elevation angel is larger than the minimum value. From Fig. \ref{fig:drl}, we can observe that the convergence of the proposed scheme can be obtained with the increase of time slots. Additionally, through comparing the converged performances, it can be found that the proposed scheme outperforms over the other two schemes in terms of the reward and throughput performances. Particularly, schemes with sequential policy updating manner have large advantages over the independent scheme since agents update policies in a temporarily stationary environment. Furthermore, the proposed scheme is superior to the single-estimation scheme since the sum advantage of agents is estimated by the UE. This results justifies the effectiveness of the collaboration designs in constructing multi-agent DRL schemes.

\begin{figure}[htb]	
	\begin{minipage}[b]{0.49\linewidth}
        \centering
        \includegraphics[scale=0.16]{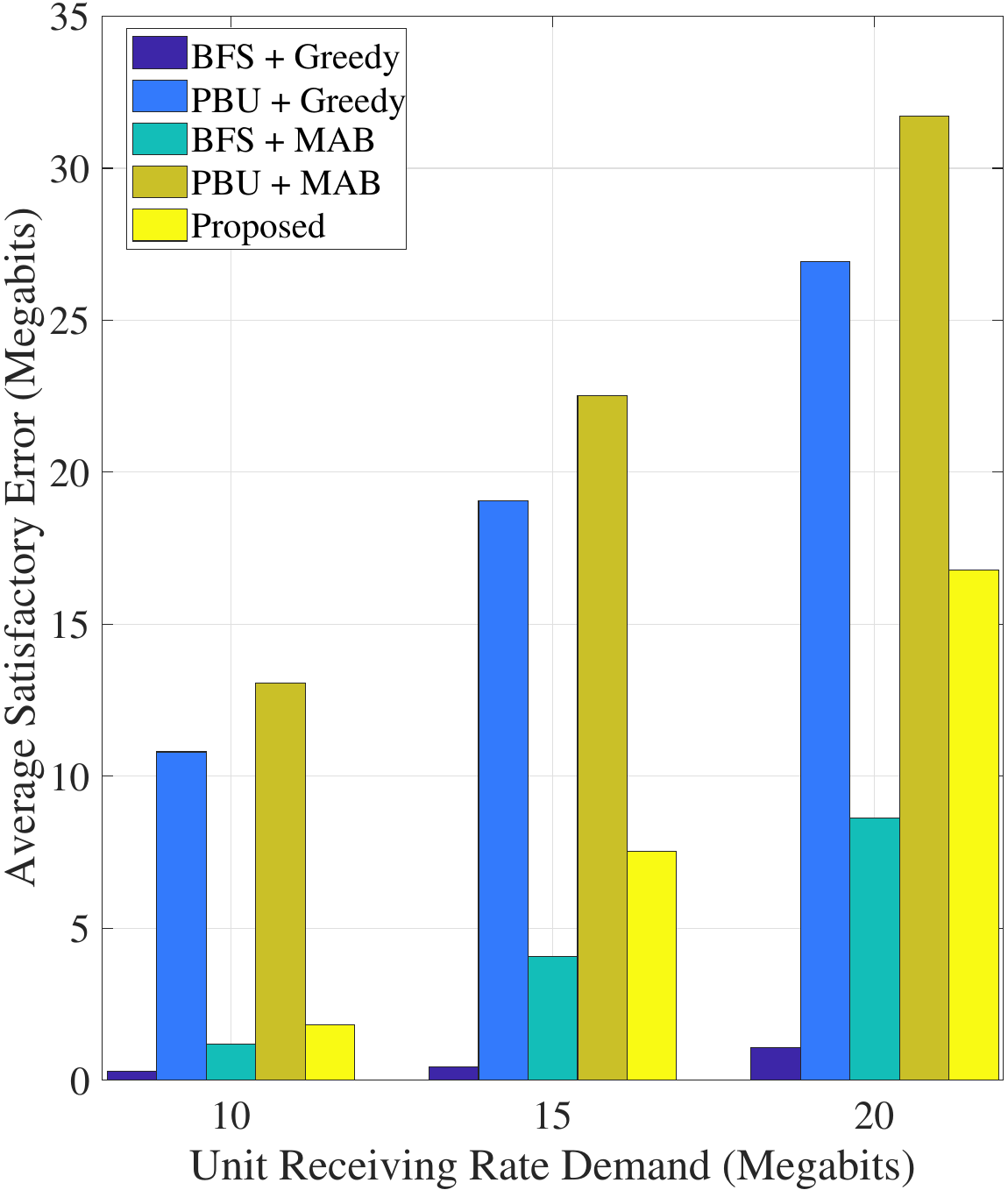}
    \end{minipage}%
    \begin{minipage}[b]{0.49\linewidth}
        \centering
        \includegraphics[scale=0.16]{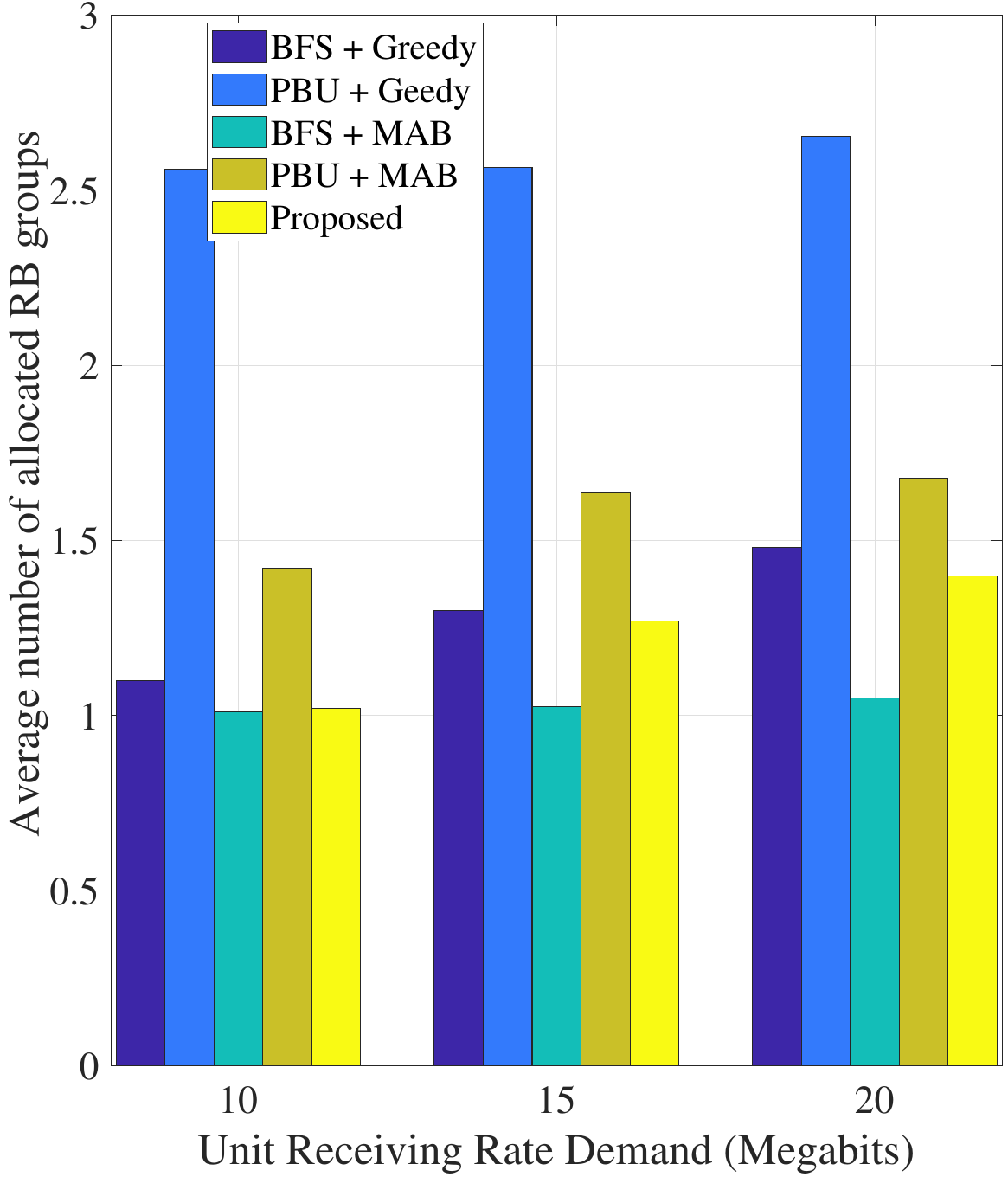}
    \end{minipage} \\[0.90mm]
    \caption{Average satisfactory error and number of allocated RB groups with different algorithms.}\label{fig:satis}
    \vspace{-0.5em}
\end{figure}

Subsequently, we present the average satisfactory error and number of allocated RB groups of the proposed scheme and separated optimization schemes in Fig. \ref{fig:satis}. In BFS-based schemes, the beam sweeping is executed every 10 degree. The satisfactory error performance is defined as the absolute value of satisfactory punishment of UE, and it is expected to approach zero. From Fig. \ref{fig:satis}, we can find that the proposed scheme can achieve a close satisfactory error performance to that of the BFS-MAB scheme, and can outperform the other two PBU-based schemes. Additionally, the number of RB groups utilized in the proposed scheme is large than the BFS-MAB scheme, and less than the other three schemes. In Table II, we adopt the weighted-sum utility defined in \cite{Senouci2016} to further evaluate the composite score of satisfactory error, number of RB groups and computational complexity of each of different schemes. Trough comparing the weighted-sum utilities of different schemes, we can find the proposed scheme can achieve the minimum value over the other four schemes under different weight combinations. Hence, an effective trade-off between computational complexity and transmission performance can be achieved by the proposed scheme.

\begin{table*}
\centering
\small
\caption{Weighted-sum Utility of Different Schemes.}
\begin{tabular}{ccccccc}
\hline
\hline
\multirow{2}{*}{\thead{Utility \\ Coefficients}}&
\multirow{2}{*}{\thead{Unit Receiving Rate Demand (Megabits)}}
&\multicolumn{5}{c}{Scheme}\cr\cline{3-7}
& & \thead{BFS-Greedy}&\thead{PBU-Greedy}&\thead{BFS-MAB}&\thead{PBU-MAB}&Proposed\cr
\hline

\multirow{3}{*}{$[\frac{1}{3},  \frac{1}{3}, \frac{1}{3}]$}
     & 10 & 0.2794 & 0.3642 & 0.2872 & 0.3152 & 0.1056\cr\cline{2-7}
     & 15 & 0.2896 & 0.3554 & 0.3027 & 0.3199 & 0.1590\cr\cline{2-7}
     & 20 & 0.3017 & 0.3493 & 0.3181 & 0.3110 & 0.2001\cr\hline
\multirow{3}{*}{$[\frac{1}{2},  \frac{1}{4}, \frac{1}{4}]$}
     & 10 & 0.3539 & 0.2732 & 0.3597 & 0.2364 & 0.0793\cr\cline{2-7}
     & 15 & 0.3615 & 0.2666 & 0.3713 & 0.2399 & 0.1193\cr\cline{2-7}
     & 20 & 0.3706 & 0.2620 & 0.3829 & 0.2333 & 0.1501\cr\hline
\multirow{3}{*}{$[\frac{1}{4},  \frac{1}{2}, \frac{1}{4}]$}
     & 10 & 0.2130 & 0.4026 & 0.2298 & 0.3930 & 0.1011\cr\cline{2-7}
     & 15 & 0.2201 & 0.3951 & 0.2545 & 0.3919 & 0.1701\cr\cline{2-7}
     & 20 & 0.2312 & 0.3859 & 0.2783 & 0.3793 & 0.2273\cr\hline
\multirow{3}{*}{$[\frac{1}{4},  \frac{1}{4}, \frac{1}{2}]$}
     & 10 & 0.2713 & 0.4168 & 0.2721 & 0.3161 & 0.1365\cr\cline{2-7}
     & 15 & 0.2871 & 0.4045 & 0.2822 & 0.3278 & 0.1875\cr\cline{2-7}
     & 20 & 0.3032 & 0.4000 & 0.2932 & 0.3205 & 0.2227\cr
\hline
\hline
\end{tabular}\label{tab:utility}
\end{table*}

\vspace{-0.15cm}
\section{Conclusion}
In this paper, we investigate the joint optimization of beam management and resource allocation in earth-fixed cell in the NTN, and propose a two-time-scale collaborative DRL to enable the LEO satellite/UE as higher/lower-tier agents to perform decision-making tasks with different control cycles. Based on the sequential policy updating manner, the UE first updates its policy based on the sum estimated policy advantage of agents. Then, based on the decision trajectory predicted by the UE, the LEO satellite updates its policies through finite-step rollout policy. With the proposed scheme, agents at different tiers can improve monotonic improvements on value functions through collaborations. Simulation results demonstrate that the proposed scheme can perform superior performance in tackling the tradeoff between the satisfactory error, number of RB groups and computational complexity over greedy-based and PBU-based schemes.

\vspace{-0.1cm}


\begin{appendices}
\section{Proof of Remark 1}
Firstly, we proof that the higher-tier value function can be improved, i.e.,
\begin{align}
&V^{\bar{\pi}_H}({\bm{s}}_H;\bar{\pi}_L) \\\nonumber
&= \max\limits_{\bar{\pi}_H }\mathbb{E}[R_H({\bm{s}}_H, \bar{\pi}_H({\bm{s}}_H); \tau_L(\bar{\pi}_L)) + \gamma_HV^{\bar{\pi}_H}(h({\bm{s}}_H, \bar{\pi}_H, \bar{\pi}_L); \\
                     & \bar{\pi}_L) ]\\\nonumber
                     & = \max\limits_{\bar{\pi}_H }\mathbb{E}[R_H({\bm{s}}_H, \bar{\pi}_H({\bm{s}}_H); \tau_L(\bar{\pi}_L)) +\gamma_HV^{\pi_H}(h({\bm{s}}_H, \bar{\pi}_H, \bar{\pi}_L); \\
                     &\pi_L) -V^{\pi_H}({\bm{s}}_H; \pi_L) + V^{\pi_H}({\bm{s}}_H; \pi_L)]\\\nonumber
                     & \ge \mathbb{E}[R_H({\bm{s}}_H, \pi_H({\bm{s}}_H); \tau_L(\bar{\pi}_L)) +\gamma_HV^{\pi_H}(h({\bm{s}}_H, \pi_H, \bar{\pi}_L); \pi_L)\\
                     & -V^{\pi_H}({\bm{s}}_H; \pi_L) + V^{\pi_H}({\bm{s}}_H; \pi_L)]\\
                     &= \max\limits_{\bar{\pi}_L}[\rho_H(\pi_H, \bar{\pi}_L) - \rho_H(\pi_H, \pi_L)]+ V^{\pi_H}({\bm{s}}_H; \pi_L)\\
                     &\ge V^{\pi_H}({\bm{s}}_H;\pi_L), \forall {\bm{s}}_H \in {\cal{S}}_H
\end{align}
Next, note that the higher-tier policy updated toward the future policy estimated by the lower-tier agent, i.e.,  $\bar{\pi}_H \approx f(\bar{\pi}_L)$, the higher-tier value function can be improved. i.e.,
\begin{align}\nonumber
& V^{\bar{\pi}_L}({\bm{s}}_L;\bar{\pi}_H)  \\\nonumber
& = \mathbb{E}[R_L({\bm{s}}_L, \bar{\pi}_L({\bm{s}}_L); \tau_H(\bar{\pi}_H)) + \gamma_LV^{\pi_L}(g({\bm{s}}_L, \bar{\pi}_L, \bar{\pi}_H)); \pi_L)\\
                     &- V^{\pi_L}(s_L;\pi_H)+V^{\pi_L}(s_L;\pi_H)] \\\nonumber
                     & \approx \mathbb{E}[R_L({\bm{s}}_L, \bar{\pi}_L({\bm{s}}_L); \tau_H(f(\bar{\pi}_L))) + \gamma_LV^{\pi_L}(g({\bm{s}}_L, \bar{\pi}_L, f(\bar{\pi}_L));\\
                     & f(\bar{\pi}_L)) - V^{\pi_L}({\bm{s}}_L;\pi_H)+V^{\pi_L}({\bm{s}}_L;\pi_H)] \\
                     &= \max\limits_{\bar{\pi}_L}[\rho_L(\bar{\pi}_L, f(\bar{\pi}_L)) - \rho_L(\pi_H, \pi_L)] + V^{\pi_L}({\bm{s}}_L;\pi_H)\\
                     &\ge  V^{\pi_L}({\bm{s}}_L;\pi_H), \forall {\bm{s}}_L \in {\cal{S}}_L,
\end{align}
where $g(\cdot)$ is the state transition function of the lower-tier agent. Therefore, monotonic policy improvements of the higher-tier and lower-tier agents can be achieved after one policy updating stage, and the proof is completed.

\end{appendices}

\end{document}